\documentclass[a4paper,11pt]{article}
\usepackage{pos}
\usepackage{siunitx}
\usepackage{comment}
\usepackage{bibspacing} 
\usepackage{lineno}
\usepackage{cleveref}
\usepackage{wrapfig}

\title{Performance Studies of Layered Water Cherenkov Detectors}
\author[a]{Pierre Billoir}

\author[b,c]{Benjamin Flaggs}
\author[c]{Ioana C. Mariș}
\author*[c]{Andrea Parenti}

\affiliation[a]{Laboratoire de Physique Nucleaire et Hautes Energies, 4 place Jussieu, 75252 Paris, France }

\affiliation[b]{Bartol Research Institute, Department of Physics and Astronomy, University of Delaware 104 The Green, Newark, DE 19716, USA}

\affiliation[c]{Université Libre de Bruxelles, Department of Physics,
Plainlaan 2, CP 230, Brussels, Belgium}

\emailAdd{bflaggs@udel.edu}
\emailAdd{ioana.maris@ulb.be}
\emailAdd{andrea.parenti@ulb.be}
\emailAdd{billoir@lpnhe.in2p3.fr}

\abstract{Next-generation air-shower detectors, such as the Global Cosmic Ray Observatory (GCOS) and the Probing Extreme PeVatron Sources (PEPS) experiment, are considering water-Cherenkov detectors as a base design. A key factor in improving the sensitivity to ultra-high-energy gamma rays and to the mass composition of ultra-high-energy cosmic rays is the ability to measure the muonic content of air showers. To address this, a layered water Cherenkov tank design has been previously proposed. The water volume of the tank is divided into two optically separated layers. The electromagnetic component of the shower is mostly absorbed in the top layer, while the bottom layer records the light produced by through-going muons. Two prototype tanks were deployed at the Pierre Auger Observatory site in 2014 and have been recording data for more than 10 years. We present the performance of the prototype tanks and compare it with simulations, focusing mostly on the calibration. We investigate different dimensions for the water volumes. For the GCOS Observatory, one important challenge is to cover extremely large surfaces of \SI{40000}{\km \squared} to \SI{60000}{ \km\squared} and achieve 100\% efficiency at \SI{10}{EeV}. Based on the size of the footprint of air-showers, we compute the number and spacing of detectors needed to fulfill the GCOS requirements.}

\ConferenceLogo{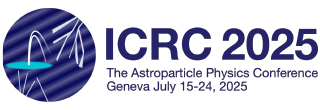}

\FullConference{39th International Cosmic Ray Conference (ICRC2025)\\
 15–24 July 2025\\
Geneva, Switzerland\\}

\begin{document}
\maketitle

\section{Introduction}
Long-standing questions such as the origin and composition of UHECRs and the nature of Galactic PeVatrons remain, to this day, unanswered \cite{UHECR_whitepaper}. The main objective of the next-generation air-shower arrays like GCOS \cite{GCOS_arxiv,GCOS_ICRC23, GCOS_ICRC25} will be to make more precise measurements of the cosmic ray flux and composition at energies above \SI{50}{EeV}. The main strategy in tackling these issues will be the deployment of detectors covering larger surface areas, with a large duty cycle, combined with the use of state-of-the-art detection techniques.

For surface detectors, the separation between electromagnetic and hadronic components of the air showers can lead to the identification of the primary particle initiating an air shower. This is fundamental to better constrain the cosmic ray origin at ultra-high energies and to increase the sensitivity to gamma rays in the PeV region.  Moreover, probing both the electromagnetic and muonic air-shower components can test the hadronic models currently in use, a main source of systematic uncertainty in the measurement of UHECRs\cite{AugerHadronic}.

One of the proposed designs for future arrays is a layered water-Cherenkov detector (LCD), originally proposed as part of the AugerPrime upgrade of the Pierre Auger Observatory \cite{LSD}. This concept was developed starting from the fixed geometry of the Auger water-Cherenkov detector. The water volume was divided into two, using optically separated, stacked layers. The electromagnetic component of the shower ($\gamma$, e$^{\pm}$) will be highly attenuated and absorbed in the top layer (attenuation length of about \SI{40}{\cm}), with mostly muons depositing energy in the bottom layer. This modification of the water Cherenkov detector allows the separation between electromagnetic and muonic components of each air shower, with muon number relative resolutions as good as 10-15\% in a single detector unit.  This design is compact and cost-effective, durable over time, and low maintenance. The Global Cosmic Ray Observatory (GCOS), Probing Extreme PeVatron Sources (PEPS) experiment, and the Southern Wide-field Gamma-ray Observatory (SWGO) \cite{PEPS, PEPS2, SWGO_WhitePaper, SWGO_LSD} are all planning the use of layered water Cherenkov detectors as the main unit of their surface detectors.

In this work, the prototype layered Cherenkov detectors deployed at the Pierre Auger Observatory site are introduced, leading to a discussion of the signal calibration method. Then, a simulation-based study is presented, investigating different values of the height of the optical volumes. Finally, the GCOS trigger efficiency above \SI{10}{EeV} as a function of the detector number and spacing will be presented. 

\section{Prototype detectors at the Pierre Auger Observatory}

\begin{figure}[t]
\centering

\begin{minipage}{.55\textwidth}
  \centering
  \vspace*{-3ex}
  \includegraphics[width=\textwidth]{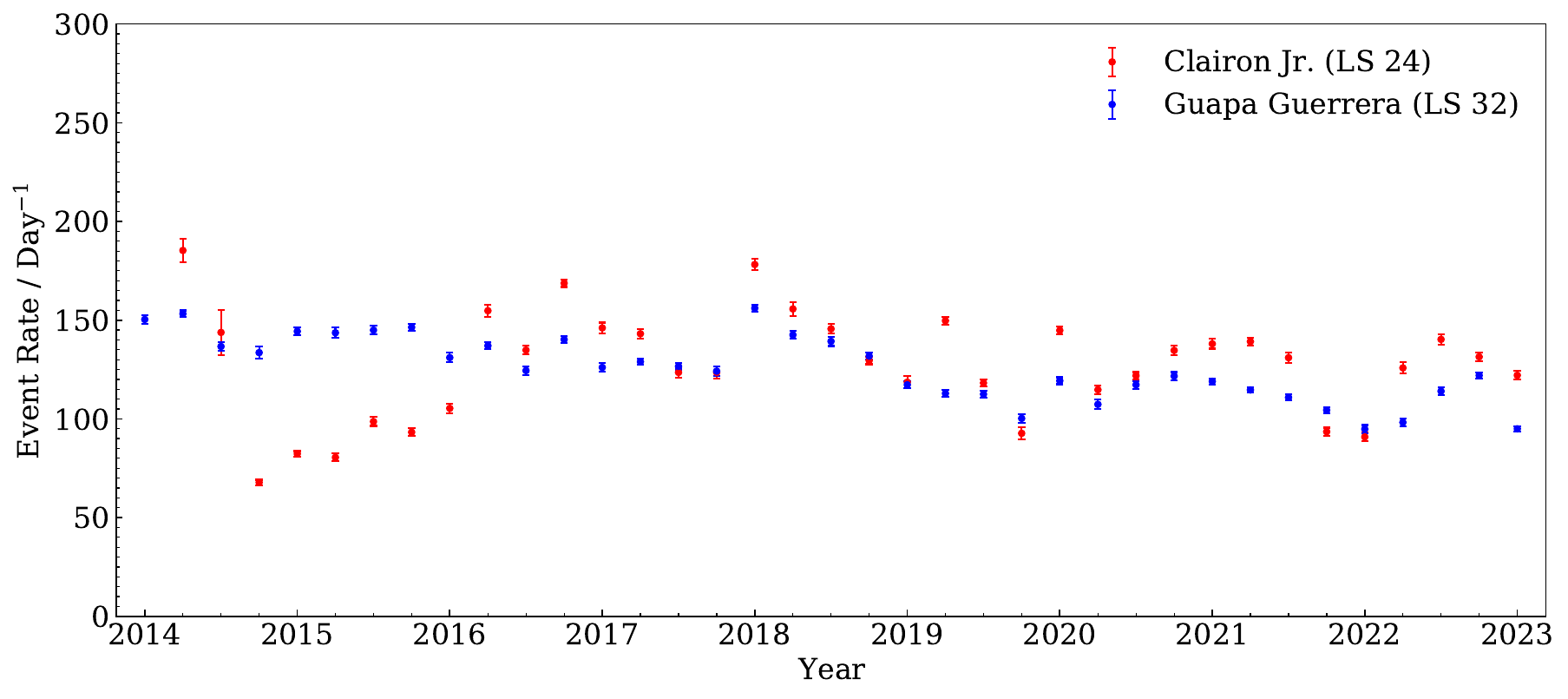}
  \vspace*{-4.5ex}
  \includegraphics[width=\textwidth]{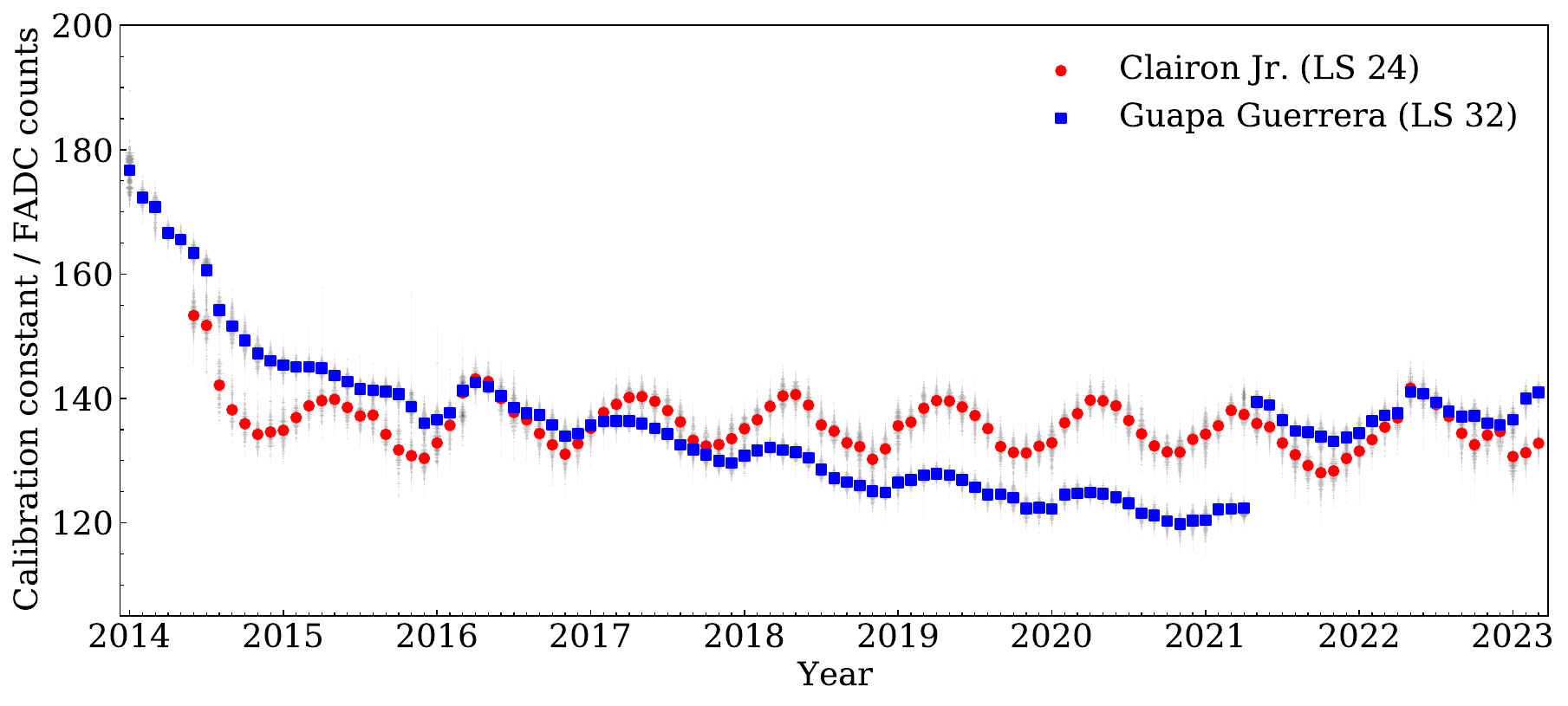}
\end{minipage}
\begin{minipage}{.4\textwidth}
  \centering
    \vspace*{-3ex}
\includegraphics[clip, width=\textwidth]{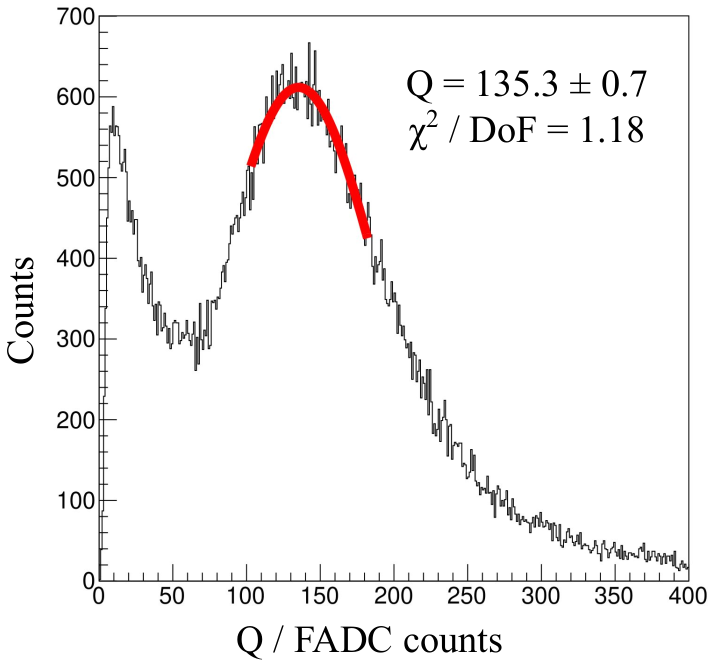}
\end{minipage}\qquad

    \caption{(left panel, top) The daily event rate for Guapa Guerrera and Clairon Jr since their installation. (left panel, bottom)  The average calibration constants for the bottom layers. (right panel) Example of a charge calibration histogram for the bottom layer.}
    \label{fig:bottom_cal}
\end{figure}

Two prototype detectors have been built and installed at the Auger site as part of the proposal for the AugerPrime upgrade in 2014~\cite{LSD}. As they were planned as modifications of the standard Auger detectors, they were bound to their geometry.
These detectors are cylindrical tanks with a diameter of \SI{3.6}{m} and a height of \SI{1.2}{m}. The light is collected by three 9-inch PMTs looking into the water volume from the top of the tank. The LCD prototypes were built using a reflective separation layer at \SI{0.8}{m} height from the bottom of the detector, separating the water volume into two. Three (or two) PMTs were kept for the top layer, while another PMT was installed to measure the light from the bottom volume of the tank.  Five detectors have been built, with two of them, Guapa Guerrera and Clairon Jr., constantly taking data since their installation in 2014\footnote{Three prototypes were dismantled by the Auger Collaboration to be used for spares of the main Auger array.}. A new look at the data has been performed in~\cite{Flaggs:2025N+} and also presented here. The event daily rate as a function of time is illustrated in \cref{fig:bottom_cal} (left panel, top). There have been no interventions during this period, besides a recalibration intervention for Clairon Jr. in 2016 and a fix of the Guapa Guerrera electronics in 2021. Both detectors were upgraded to the new Auger electronics in April 2023. 

The electromagnetic and muonic signals can be extracted from the signal collected in the top and bottom layers from a set of linear equations:

\begin{equation}
    \begin{split}
        & S_{\textrm{top}}  = a \cdot S_{\gamma,e^\pm} + b \cdot S_{\mu} \\ 
        & S_{\textrm{bottom}} = (1-a) \cdot S_{\gamma,e^\pm} + (1-b) \cdot S_{\mu} 
    \end{split}
\end{equation}

where $a$ and $b$ depend mainly on the detector geometry. The coefficients were estimated with simulation to be $a \approx 0.6$ and $b \approx 0.4$ in ~\cite{LSD}. Notably,  they are independent of the UHECR primary particle, energy, direction (up to a zenith angle of 60$^\textrm{o}$), and hadronic interaction model. Therefore, by solving the equations, $S_{\gamma, e^\pm}$ and $S_{\mu}$ can be extracted on a station-by-station basis from the calibrated top and bottom signals. 

\subsection{Calibration}

The standard calibration procedure for water-Cherenkov detectors \cite{SD_calibration} is a conversion of the output signal, in FADC counts, to 
Vertical Equivalent Muon (VEM) units. One VEM is the charge deposited by a muon crossing the center of the tank vertically, defining a unit that can be used uniformly across multiple detectors. The calibration is performed by collecting the light from omnidirectional atmospheric particles produced in low-energy air showers with a well-defined constant for the conversion to VEM. A tank records a signal when at least one PMT signal is above a certain threshold (for more details, see \cite{SD_calibration}). The signals are integrated over time in the local station software to obtain the deposited charge and afterwards filled into a histogram. The histogram, with an example for the bottom layer illustrated in \cref{fig:bottom_cal} (right panel), exhibits two main features: a peak at around 135 FADC counts corresponding to atmospheric muons and an exponentially decreasing population at lower FADC counts due to the electromagnetic particles in the air showers. The position of the muon peak is used to define the FADC value of the VEM unit.

In the bottom layer, a large fraction of the particles triggering the PMTs will be muons, because the electromagnetic ones are highly attenuated in the top layer. For this reason, the muon peak is very well distinguishable in the charge distribution, allowing for a clean and precise calibration. The calibration histograms, constructed with the signals from one-minute intervals, are recorded for each event to account for variation due to seasonal atmospheric modulation and temperature dependence of the detector response. These effects can be seen in \cref{fig:bottom_cal} (left panel, bottom), where the average calibration constants for Guapa Guerrera and Clairon Jr. exhibit the expected seasonal modulation. The decrease observed in the beginning ceases after a few months, necessary for the water in the tank to settle. For Guapa Guerrera, stable calibration was achieved in 2021, when the electronics were replaced. Clairon Jr presents a stable calibration in the full time period. 

For the top layer PMTs, this calibration method is more challenging: due to the shorter track length of the particles inside the volume, the energy deposits by electromagnetic particles and muons tend to overlap and the tail of the charge deposit by electromagnetic particles partially hides the muon peak used for calibration. A coincidence between the bottom PMT signal and the top PMT will allow for a large reduction of the electromagnetic component. The implementation of this trigger condition is a subject of further studies.

\section{Simulation study of calibration histograms}

\begin{figure}[]
    \centering
    \includegraphics[width=0.48\linewidth]{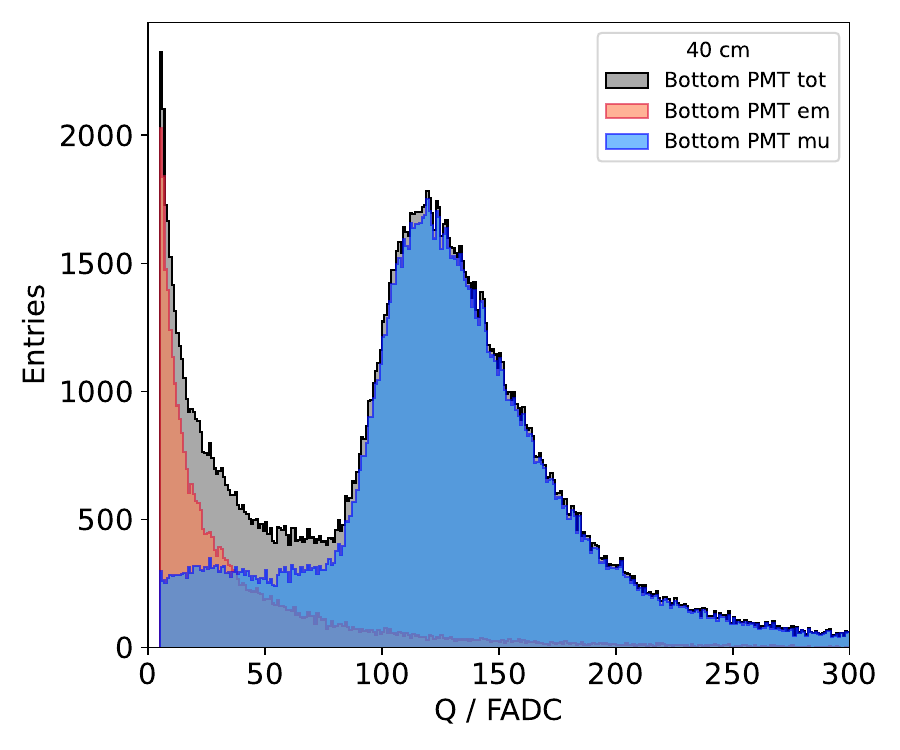}\hfill\includegraphics[width=0.48\linewidth]{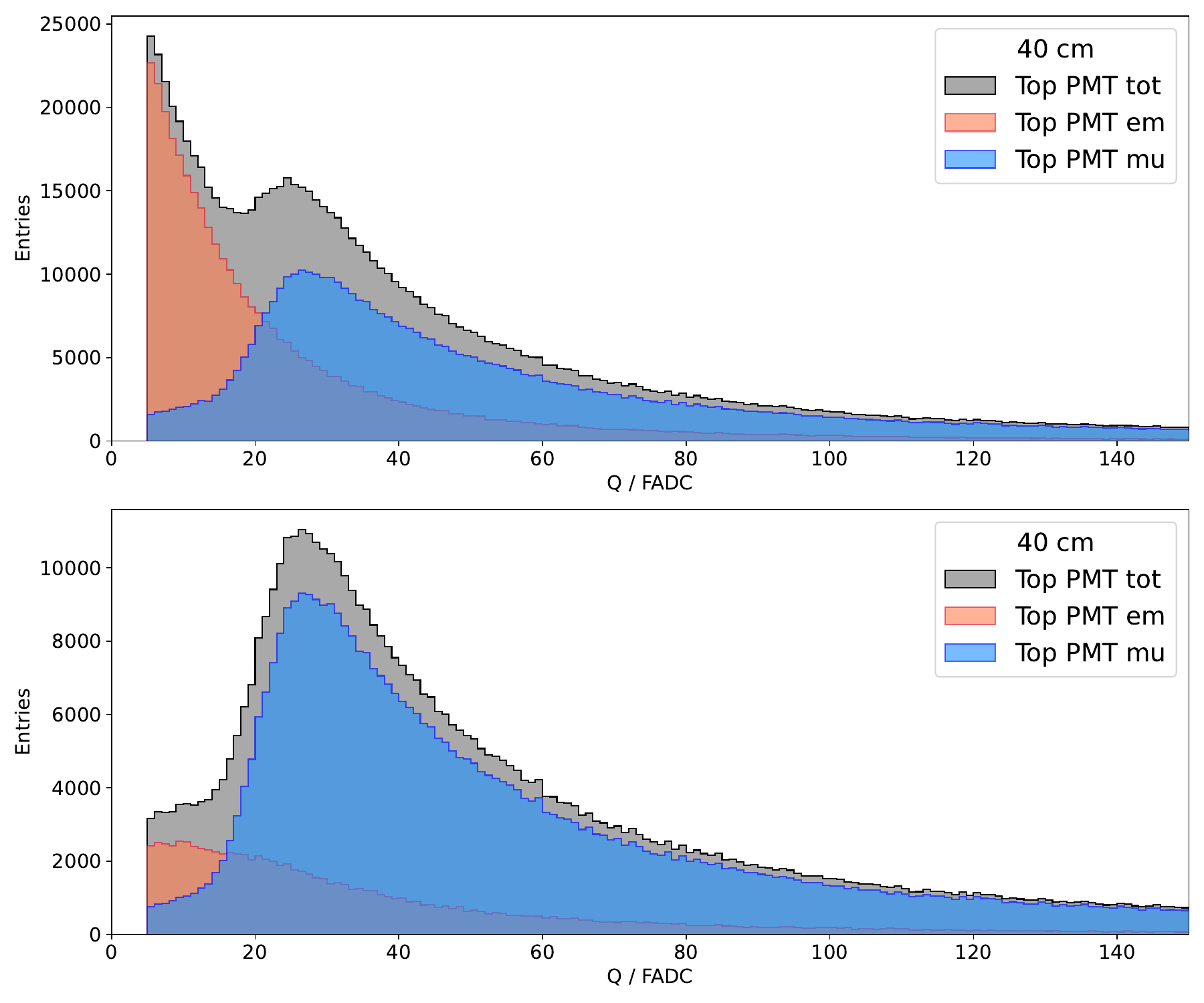}
    \caption{(left panel) The simulated charge from the bottom PMT of the double liner tank with top height of 40 cm. (right panel, top) Simulated charge from the top PMT. (right panel, bottom) Charge histogram filled with events that satisfy a request of coincidence between top and bottom PMT.}
    \label{fig:fadc_top_bottom}
\end{figure}

To reproduce the calibration histograms we performed simulations of the layered tank response to atmospheric particles. A stand-alone software was used where the most important interactions of the electromagnetic particles and of the muons within the water volume have been included. The production of Cherenkov photons, with their optical propagation, reflection, and absorption has been simulated taking into account the tank properties. Finally, the PMTs quantum efficiencies and the FADC sampling have been simulated. The simulation code is much faster than a full Geant4 implementation, being optimised by computing the number of reflections off the walls and the expected path and not following every generated photon, significantly speeding up the computing time. Via its configuration files, it easily allows for changes in the tank geometry, optical properties, and position of the PMTs. 

\begin{wrapfigure}{r}{0.5\textwidth}
  \begin{center}
    \includegraphics[width=0.46\textwidth]{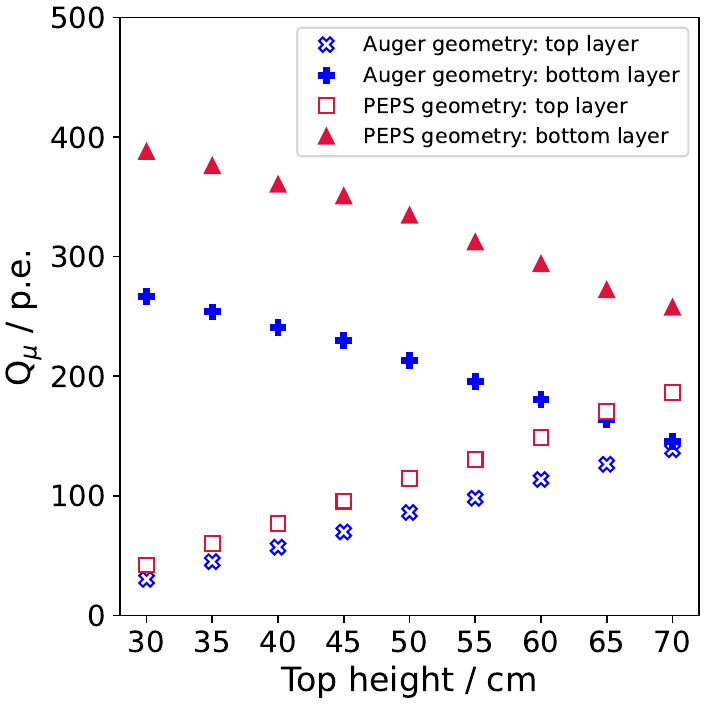}
  \end{center}
  \caption{Number of the muon peak photoelectrons for the top and bottom layer of a simulated tank with the Auger prototypes geometry (blue markers, total height of \SI{1.2}{\m}, diameter of \SI{3.6}{\m}) and the PEPS tank geometry (red markers, total height of \SI{1.3}{\m}, diameter of \SI{3}{\m}).}
  \label{fig:pe_vs_height}
\end{wrapfigure}

To simulate omnidirectional particles, a collection of CORSIKA~\cite{corsika} showers was used, in a rigidity range from \SI{10}{GV} to \SI{10}{TV}, of mostly p, He primaries (98\%) with a minor contribution from heavier nuclei O, Si, Fe. The properties of the secondary particles that reach the ground at the Pierre Auger Observatory altitude are recorded, including their energy and direction.  We simulated different top-layer heights between \SI{30}{\cm} and \SI{70}{cm}, keeping the total tank height fixed at \SI{120}{cm} as a first approach. For each different value of the top layer height, a sample of $10^6$ particles is randomly drawn from the stored collection and simulated.

The deployed stations have, as previously mentioned, two layers of \SI{40}{\cm} and \SI{80}{\cm}. In \cref{fig:fadc_top_bottom} the results of simulation are shown. The charge histograms reproduce the measurements taken by the Auger prototype \cref{fig:bottom_cal}(right panel) for the bottom layer, validating the simulations. The small difference in the position of the muon peak can be explained by a simple difference in the PMT gain between data and simulation. The average charge collected by the top PMTs is shown in the same figure. In the top layer, the muon hump is partially swallowed by the electromagnetic background (right panel, top). By requesting a coincidence between the top and the bottom PMT, the muon peak prominence increases significantly and makes it usable for an accurate calibration. As expected, the request of a coincidence removes a large part of the electromagnetic component.

The time response of the PMT is also simulated: the average waveform for each PMT is computed and fitted with an exponential distribution. For a 40 cm top layer, the time decay constant of the top layer signal is $\tau \sim $\SI{22}{ns}, while for the bottom layer it is \SI{\sim 39}{ns}.

Different heights are studied as well as different diameters. For PEPS, the base design will increase the size of the top layer to improve the separation of the air-shower components, and the diameter reduced to \SI{3}{m} to ease the transport and the installation. The results of these scans are shown in \cref{fig:pe_vs_height}. As can be seen, the number of photoelectrons increases with reducing the size of the diameter due to an increase in the number of reflections and a smaller absorption in the water volume. In terms of time response, the time decay constant for the PEPS design is similar to the Auger one, with $\tau \sim $\SI{24}{ns} for the top layer and $\tau \sim$\SI{38}{ns} for the bottom layer.
These represent the first studies of the LCD in the context of GCOS and PEPS and are the basis for further performance studies involving full air-shower simulations.  

\section{Required number of detectors for GCOS}

\begin{figure}
    \centering
    \includegraphics[width=0.42\linewidth]{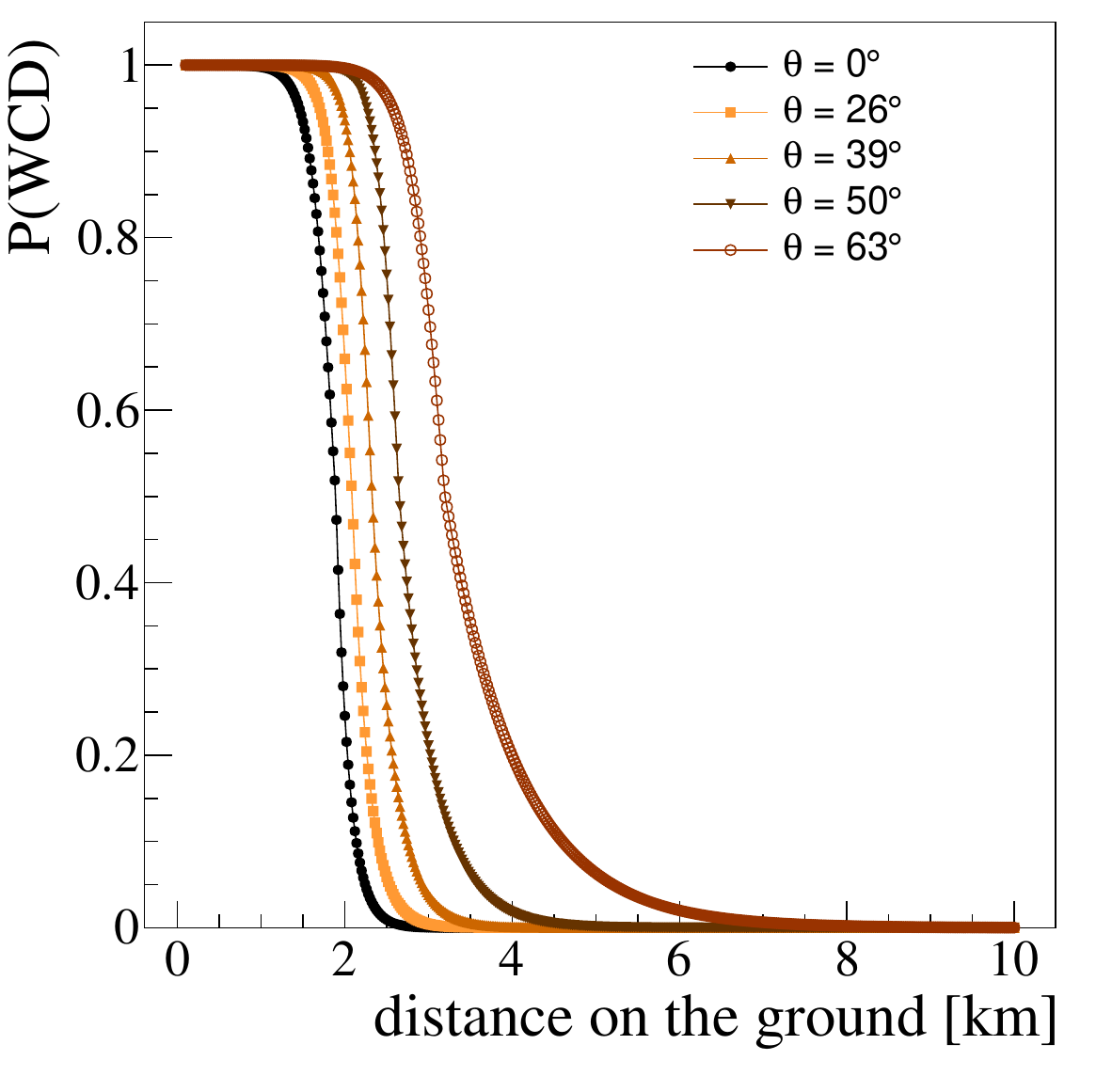}
    \includegraphics[width=0.45\linewidth]{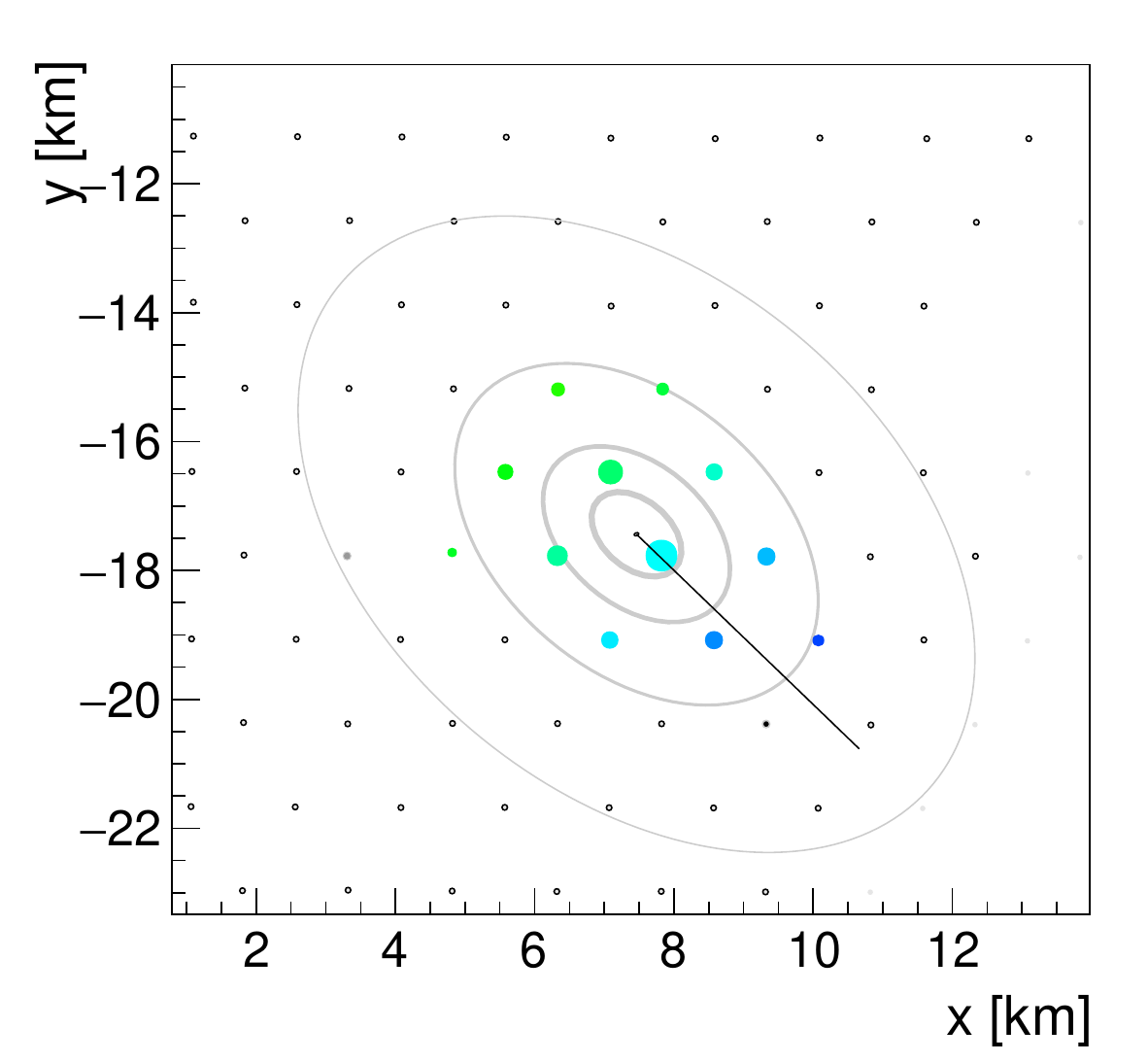}\\
    \includegraphics[width=0.42\linewidth]{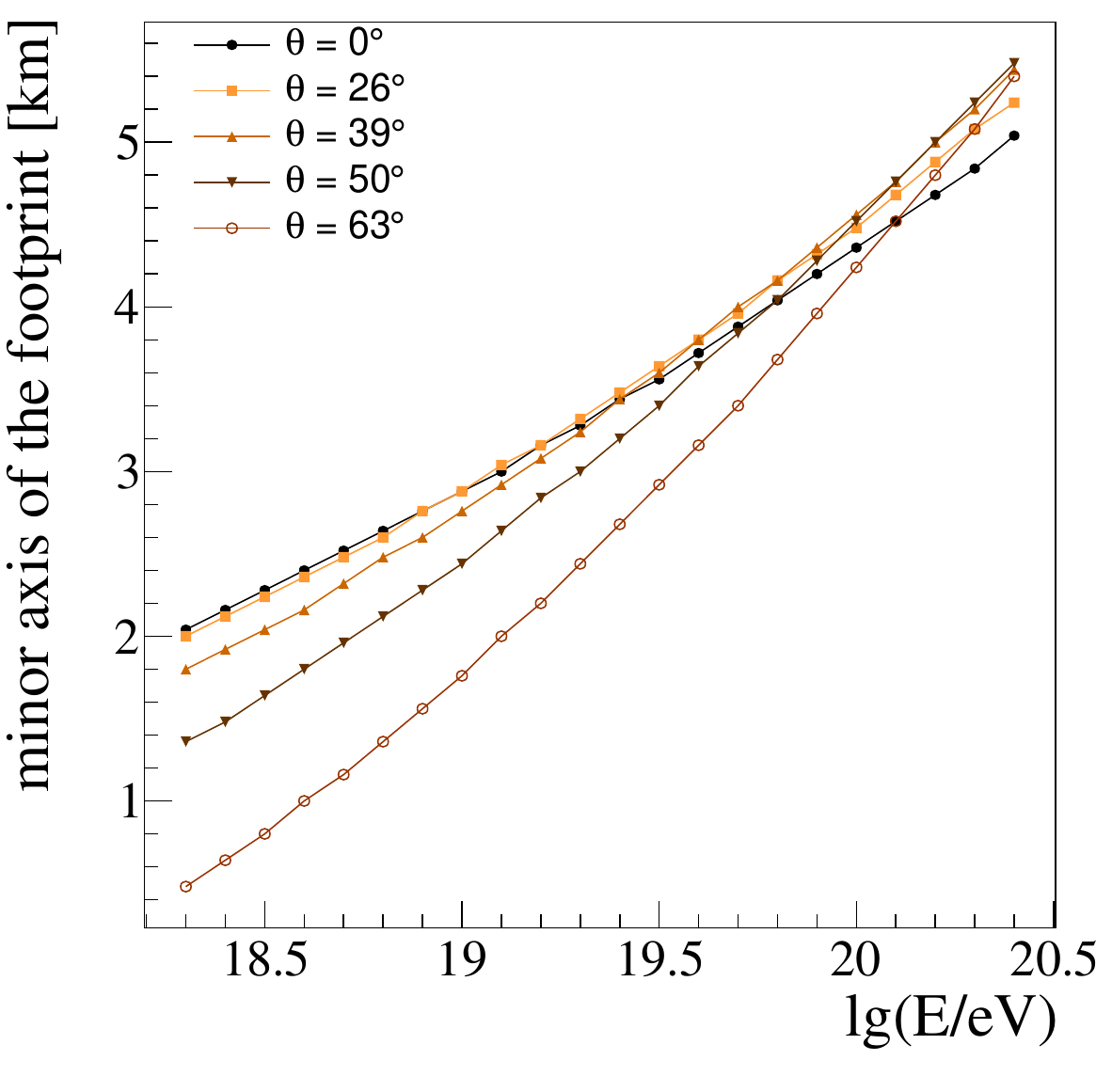}
    \includegraphics[width=0.42\linewidth]{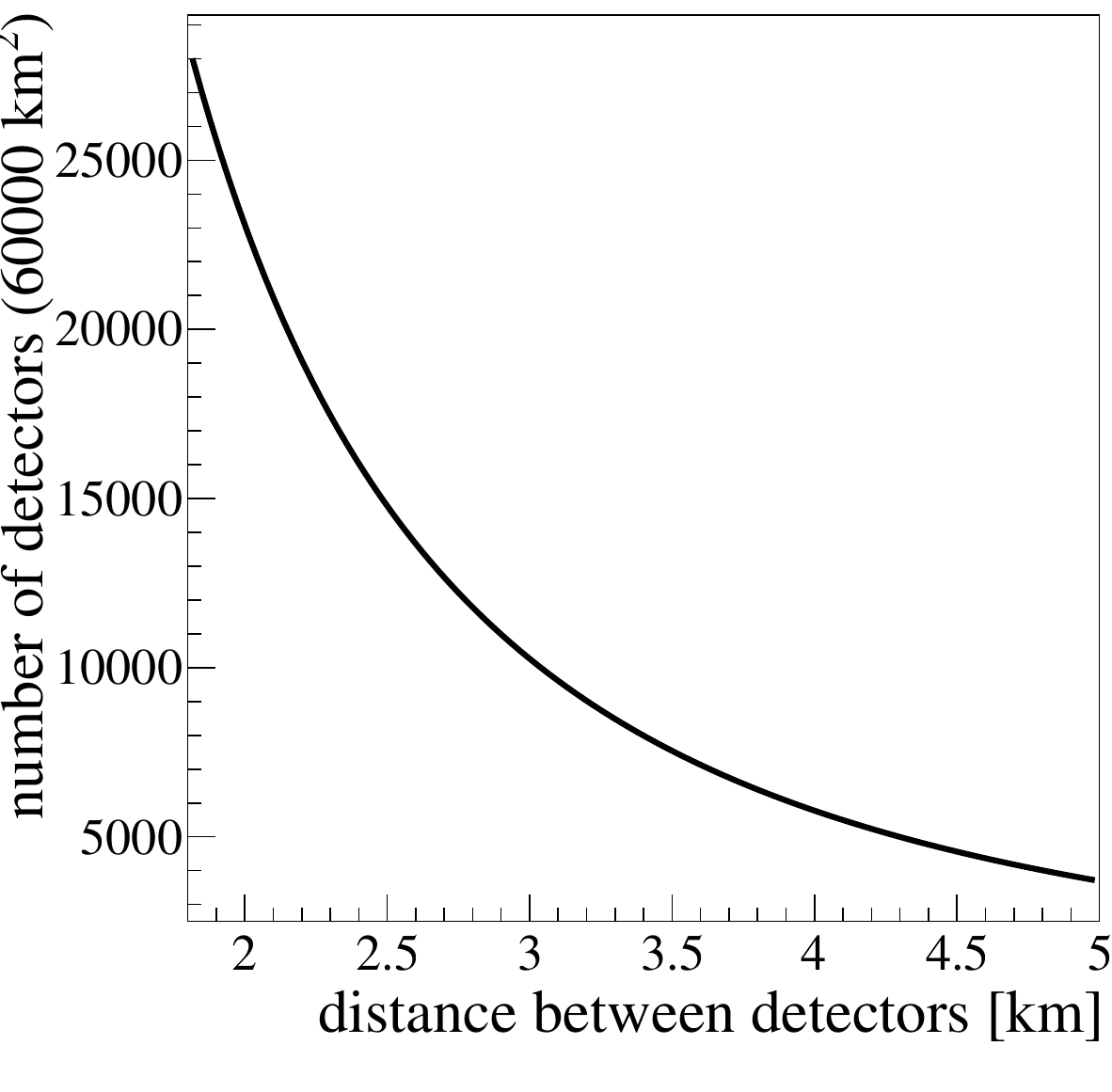}
    \caption{ (left panel, top) Lateral trigger probability of a single tank for proton-induced showers at lg(E/eV) = 19.  (right panel, top) Example of an event at 10 EeV and 48 degrees zenith angle. (left panel, bottom) Example of the size of the footprint on the ground for iron-initiated showers; the axes of the ellipse are defined at 90\% station trigger efficiency. (right panel, bottom) Number of detectors necessary to cover a surface of \SI{60 000}{km^2} as a function of the distance between the tanks, for a triangular grid array.}
    \label{fig:GCOS}
\end{figure}

The current strawman design of GCOS assumes an area of \SI{60 000}{km^2} to be covered with surface detectors to allow one to accumulate enough statistics at the highest energies. The Observatory features an air-shower array to sample the density of electromagnetic particles and muons at ground and a fluorescence and/or a radio detector to provide the energy and mass scale of the array~\cite{GCOS_arxiv}. The strawman design assumes a 100\% trigger efficiency above \SI{10}{EeV}. This allows for high-quality data (multiplicity of  trigger detectors larger than 5) to be collected starting from \SI{30}{EeV}.

Assuming water-Cherenkov detectors, with a similar size to the Pierre Auger Observatory ones, and air-shower simulation of a mixed cosmic rays composition, we define the lateral trigger probability for a simple time-over-threshold trigger for individual stations. The result for proton induced showers with energies between \SI{10}{\exa\eV} and \SI{100}{\exa\eV} and different zenith angles are shown in \cref{fig:GCOS}.
The station trigger probability as a function of the distance to the shower core, shows the expected behavior with a full efficiency of up to \SI{2}{\km} from the air-shower axis. The probability rapidly decreases for distances larger than about \SI{3}{\km}, depending on the shower inclination. A footprint of a shower at \SI{10}{EeV} and 48 degrees is shown in the same figure. Looking at the shower footprint, the smallest distance on the ground is defined as the minor axis of the ellipse at 90\% single station trigger probability, from which we obtain the maximum spacing that we can afford between the detectors. With about \SI{2.2}{\km} spacing almost all air showers above \SI{10}{EeV} will trigger the array. Relaxing the distance requirement would have a strong influence on the trigger probability: a distance of \SI{3}{\km} would make the array fully efficient just at about \SI{30}{\exa\eV}.

Assuming a triangular grid arrangement, the total number of detectors necessary to cover the target area of \SI{60 000}{\km\squared} is computed as a function of the detector spacing and shown in \cref{fig:GCOS}. With a spacing below \SI{2.5}{km}, more than \qty{15 000} detectors are needed. The large number of detectors necessary for GCOS poses a difficult challenge, which will require failure- and maintenance-free detectors, including an industrial-scale production. 
\section{Conclusion and outlook}
With next-generation surface shower arrays on the horizon, new detector technologies are currently being developed. The layered water Cherenkov tank is a promising candidate for future experiments such as PEPS, GCOS and SWGO, offering a durable instrument with the capability of measuring separately the electromagnetic and muonic components of air showers, event-by-event. The study of such a detector is being carried out both from the experimental and the simulation side. In this work, results for the calibration of the prototype tanks at the Pierre Auger Observatory have been presented, highlighting the bottom PMT calibration method with atmospheric muons. The simulation of the layered tank response to secondary air-shower particles shows very good agreement with the measurements from the prototype. The simulation has also been used to study the response for different heights of the top layer of the tank. Moreover, the results of the simulations validate the coincidence method as a viable way to perform calibration of the top PMT. This coincidence calibration will be implemented and tested in the near future in the field.

Finally, the calculations of the number of required number of detectors for the GCOS array highlight the challenges that will come with building such a large observatory: to have a 100\% efficiency above \SI{10}{EeV}, the maximum distance allowed between tanks is \SI{2.2}{km}. Assuming such distance and a triangular grid, more than \qty{15000} tanks are needed to cover the target surface. To meet such ambitious requirements, a reliable and maintenance-free detector is necessary: the layered water Cherenkov tank is a promising option.

\bibliographystyle{JHEP}
\bibliography{biblio}

\providecommand{\href}[2]{#2}\begingroup\raggedright\begin{thebibliography}{10}

\bibitem{UHECR_whitepaper}
A.~Coleman et~al., \emph{{Ultra high energy cosmic rays The intersection of the Cosmic and Energy Frontiers}}, \href{https://doi.org/10.1016/j.astropartphys.2023.102819}{\emph{Astropart. Phys.} {\bfseries 149} (2023) 102819} [\href{https://arxiv.org/abs/2205.05845}{{\ttfamily 2205.05845}}].

\bibitem{GCOS_arxiv}
M.~Ahlers et~al., \emph{{Ideas and Requirements for the Global Cosmic-Ray Observatory (GCOS)}},  \href{https://arxiv.org/abs/2502.05657}{{\ttfamily 2502.05657}}.

\bibitem{GCOS_ICRC23}
{\scshape GCOS} collaboration, \emph{{Science with the Global Cosmic-ray Observatory (GCOS)}}, \href{https://doi.org/10.22323/1.444.0281}{\emph{PoS} {\bfseries ICRC2023} (2023) 281} [\href{https://arxiv.org/abs/2309.17324}{{\ttfamily 2309.17324}}].

\bibitem{GCOS_ICRC25}
T.~Fujii, \emph{{The Global Cosmic Ray Observatory – Challenging next-generation multi-messenger astronomy with interdisciplinary research}}, {\emph{these proceedings} (2025) }.

\bibitem{AugerHadronic}
{\scshape Pierre Auger} collaboration, \emph{{Testing hadronic-model predictions of depth of maximum of air-shower profiles and ground-particle signals using hybrid data of the Pierre Auger Observatory}}, \href{https://doi.org/10.1103/PhysRevD.109.102001}{\emph{Phys. Rev. D} {\bfseries 109} (2024) 102001} [\href{https://arxiv.org/abs/2401.10740}{{\ttfamily 2401.10740}}].

\bibitem{LSD}
A.~Letessier-Selvon, P.~Billoir, M.~Blanco, I.C.~Mari\c{s} and M.~Settimo, \emph{{Layered water Cherenkov detector for the study of ultra high energy cosmic rays}}, \href{https://doi.org/10.1016/j.nima.2014.08.029}{\emph{Nucl. Instrum. Meth. A} {\bfseries 767} (2014) 41} [\href{https://arxiv.org/abs/1405.5699}{{\ttfamily 1405.5699}}].

\bibitem{PEPS}
I.C.~Maris and N.M.~Gonzalez, \emph{{On the possibility to measure galactic photons at the altitude of the Pierre Auger Observatory}}, \href{https://doi.org/10.22323/1.444.0718}{\emph{PoS} {\bfseries ICRC2023} (2023) 718}.

\bibitem{PEPS2}
I.C.~Maris, \emph{{Probing Extreme PeVatron Sources}}, {\emph{{these proceedings}} (2025) }.

\bibitem{SWGO_WhitePaper}
{\scshape SWGO} collaboration, \emph{{Science Prospects for the Southern Wide-field Gamma-ray Observatory: SWGO}},  \href{https://arxiv.org/abs/2506.01786}{{\ttfamily 2506.01786}}.

\bibitem{SWGO_LSD}
S.~Kunwar, H.~Goksu, J.~Hinton, H.~Schoorlemmer, A.~Smith, W.~Hofmann et~al., \emph{{A double-layered Water Cherenkov Detector array for Gamma-ray astronomy}}, \href{https://doi.org/10.1016/j.nima.2023.168138}{\emph{Nucl. Instrum. Meth. A} {\bfseries 1050} (2023) 168138} [\href{https://arxiv.org/abs/2209.09305}{{\ttfamily 2209.09305}}].

\bibitem{Flaggs:2025N+}
B.~Flaggs and I.C.~Maris, \emph{{Layered Water Cherenkov Detectors for Next Generation Air-Shower Arrays}}, \href{https://doi.org/10.22323/1.484.0086}{\emph{PoS} {\bfseries UHECR2024} (2025) 086}.

\bibitem{SD_calibration}
{\scshape Pierre Auger} collaboration, \emph{{Calibration of the surface array of the Pierre Auger Observatory}}, \href{https://doi.org/10.1016/j.nima.2006.07.066}{\emph{Nucl. Instrum. Meth. A} {\bfseries 568} (2006) 839} [\href{https://arxiv.org/abs/2102.01656}{{\ttfamily 2102.01656}}].

\bibitem{corsika}
D.~Heck et~al., \emph{{CORSIKA: A Monte Carlo code to simulate extensive air showers}}, {\emph{Forschungszentrum Karlsruhe Report FZKA-6019} (1998) }.

\end{thebibliography}\endgroup

\end{document}